\documentclass[review]{elsarticle}

\usepackage{lineno}
\usepackage[unicode]{hyperref}
\usepackage{lscape}
\usepackage{geometry}
\modulolinenumbers[1]

\journal{Icarus}

\biboptions{authoryear}

\begin{document}

\begin{frontmatter}

\title{The effect of collisional erosion on the composition of Earth-analog planets in Grand Tack models: Implications for the formation of the Earth}

\author[mymainaddress,ad2]{L. Allibert\corref{mycorrespondingauthor}}
\cortext[mycorrespondingauthor]{Corresponding author}
\ead{laetitia.allibert@mfn.berlin}

\author[mymainaddress,mylastaddress]{J. Siebert}
\author[mymainaddress]{S. Charnoz}
\author[mysecondaryaddress]{S.A. Jacobson}
\author[mythirdaddress]{S.N. Raymond}

\address[mymainaddress]{Institut de Physique du Globe de Paris, Universit\'e de Paris, 1 Rue Jussieu, Paris, France}
\address[ad2]{Museum f\"ur Naturkunde, Invalidenstrasse 43, Berlin}
\address[mysecondaryaddress]{Michigan State University, Earth and Environmental Sciences, 288 Farm Ln, East Lansing, MI 48824, USA}
\address[mythirdaddress]{Laboratoire d'Astrophysique de Bordeaux, All\'ee Geoffroy St Hilaire, Bordeaux, France}
\address[mylastaddress]{Institut Universitaire de France}




\begin{abstract}
Impact-induced erosion of the Earth's early crust during accretion of terrestrial bodies can significantly modify the primordial chemical composition of the Bulk Silicate Earth (BSE, that is, the composition of the crust added to the present-day mantle). In particular, it can be particularly efficient in altering the abundances of elements having a strong affinity for silicate melts (i.e. incompatible elements) as the early differentiated crust was preferentially enriched in those. Here, we further develop an erosion model (EROD) to quantify the effects of collisional erosion on the final composition of the BSE. Results are compared to the present-day BSE composition models and constraints on Earth's accretion processes are provided. The evolution of the BSE chemical composition resulting from crustal stripping is computed for entire accretion histories of about 50 Earth analogs in the context of the Grand Tack model. The chosen chemical elements span a wide range of  incompatibility degrees. We find that a maximum loss of 40wt\% can be expected for the most incompatible lithophile elements such as Rb, Th or U in the BSE when the crust is formed from low partial melting rates. Accordingly, depending on both the exact nature of the crust-forming processes during accretion and the accretion history itself, Refractory Lithophile Elements (RLE) may not be in chondritic relative proportions in the BSE. In that case, current BSE estimates may need to be corrected as a function of the geochemical incompatibility of these elements. Alternatively, if RLE are indeed in chondritic relative proportions in the BSE, accretion scenarios that are efficient in affecting the BSE chemical composition should be questioned.
\end{abstract}


\end{frontmatter}

\section{Introduction}

The chemical composition of the Earth and other rocky bodies of the inner solar system is inherited from both nebular and post-nebular processes. Traditionally, the compositional gradient of the Solar System's bodies is essentially explained from the sequence of condensation of elements at nebular conditions during the earliest stages of solar system formation \citep{mcdonough1995composition}. However, post-nebular processes (e.g. evaporation, differentiation, collisional erosion) might be responsible for further chemical fractionation among planetary bodies \citep{boyet2005142nd, o2008collisional, siebert2012metal}. Those are thought to occur during the last stages of planetary formation, including the so-called giant impacts stage starting 2-3 Myr after CAIs and lasting up to 200Myr. Thus, this last stage of planetary formation starts slightly before the disappearance of the gas in the disk, if the gas disappears in about 3Myr \citep{jacobson2014highly, haisch2001disk}). For instance, collisional erosion can eventually lead to substantial fractionation among terrestrial bodies at different stages, from planetesimal-like bodies to embryos and planet-like bodies \citep[e.g.][]{o2008collisional, boujibar2015cosmochemical, carter2018collisional}. There are various evidences of collisional erosion during planetary accretion. These evidences can be found on various bodies of various sizes in the solar system. Among those, the relatively large core of Mercury compared to its silicate mantle has been suggested to be resulting from large mantle stripping during a giant impact \citep[e.g.][]{benz1988collisional}. Martian meteorites are also strong evidences of collisional erosion only by their existence. This is also the case of the HED group of asteroids that are thought to come from the asteroid Vesta \citep[e.g.][]{mccord1970asteroid,consolmagno1977composition, mcsween2013dawn}, such as 20 small asteroids in the surroundings of Vesta that share a similar spectral reflectivity \citep[e.g.][]{binzel1993chips}. The similarities between the Moon and Earth-mantle isotopic compositions suggest that the Moon itself results from a giant collision on the growing proto-Earth. Additionally, largely impacted planetary surfaces of the Moon, Mars or Mercury also show the potential of small impacts to displace and remove material.  Previous studies have demonstrated the potential of collisions to alter the bulk compositions of planetary bodies \citep{o2008collisional, campbell2012evidence, bonsor2015collisional, carter2015compositional, carter2018collisional, allibert2021quantitative}, especially from the erosion of an early proto-crust \citep{o2008collisional, boujibar2015cosmochemical, carter2018collisional}. An early crust is likely formed on growing rocky bodies during accretion \citep{kleine2004182hf, elkins2011chondrites} and is assumed to be enriched in incompatible elements (elements that have an affinity for silicate melts). Consequently, proto-Earth crustal stripping can imply significant fractionation among chemical elements proportionally to their degree of incompatibility \citep{o2008collisional, boujibar2015cosmochemical, carter2018collisional, allibert2021quantitative}. For instance, this mechanism has been proposed to explain the observed superchondritic $^{142}$Nd/$^{144}$Nd through a 6\% enrichment of Sm compared to Nd in the BSE as Sm is less incompatible than Nd. Because $^{142}$Nd is the radioactive decay product of the $^{146}$Sm, that is a radioactive system characterized by a short half-life \citep{meissner1987half, kinoshita2012shorter}, this would require early erosion (within the first 30 million years) not to significantly alter the $
^{147}$Sm-$^{143}$Nd long-lived system. However, this interpretation has recently been questioned \citep{burkhardt2016nucleosynthetic, bouvier2016primitive, boyet2018enstatite}.  An alternative hypothesis proposes that nucleosynthetic anomalies observed among terrestrial precursors (i.e. chondrites) \citep{burkhardt2016nucleosynthetic, bouvier2016primitive, boyet2018enstatite} could explain the $^{142}$Nd composition of the Earth and would require very limited impact-induced RLEs fractionation, especially at early stages (i.e. before 30 Myr). However, a more recent study has determined the precise contribution of nucleosynthetic anomalies to the BSE composition in $^{142}$Nd \citep[][]{frossard2022earth}. They predict that among the initially predicted 20-ppm excess, even after considering nucleosynthetic anomalies in chondritic material, a 7.9-ppm  $^{142}$Nd excess is still required in the BSE compared to chondrites.  They argue that this remaining difference can be explained by collisional erosion. \newline
Given these competing perspectives, the process of collisional erosion and its efficiency on the fractionation of the BSE composition have to be further explored. Notably, this is a crucial question as current BSE models are built from the assumption that the RLEs abundances measured in the most fertile mantle rocks are chondritic. \newline

In this work, we quantitatively  model the effect of collisional erosion on the final composition of the BSE. A series of given chemical elements with varying geochemical properties including various degrees of incompatibility is looked into. This approach can either challenge the common assumption of chondritic proportions of RLEs in the BSE or better evaluate the conditions of terrestrial planets accretion that keep this assumption untouched. The predicted depletion of lithophile elements needs to be in reasonable agreement with geochemical models. This means that it is necessary to compare the outcomes of different collisional histories in terms of the fractionation (after collisional erosion) of RLEs with estimates from the BSE composition, which will bound the efficiency of collisional erosion during accretion. \newline

 The collisions experienced by the growing differentiated embryos during the last stages of planetary accretion can lead to preferential loss of material from chemically distinct layers (i.e. crust vs. mantle or mantle vs. core) \citep[e.g.][]{svetsov2011cratering, carter2015compositional,  shibaike2016excavation, carter2018collisional, allibert2021quantitative}. The amount of eroded material depends on the impact parameters (e.g. velocity, angle, mass ratio between impactor and target) which are linked to the dynamics of the disk  \citep{leinhardt2011collisions, bonsor2015collisional, carter2015compositional, carter2018collisional}. The efficiency of crustal stripping has been previously investigated using either an erosion model focused on the period of the Late Heavy Bombardment \citep{shibaike2016excavation}, the timing of which is in question \citep{zellner2017cataclysm, morbidelli2018timeline, mojzsis2019onset}, or from a model neglecting the erosive potential of small planetesimals \citep{carter2018collisional}. Notably, the effect of collisional erosion on the Sm/Nd ratio of the Earth was performed, in a viable dynamical context, using erosion scaling laws from smooth particle hydrodynamics (SPH) simulations of giant impacts and post-processing N-body numerical simulations of Earth's accretion \citep{carter2018collisional}. However, SPH numerical simulations do not provide a sufficiently high resolution (i.e. number of particles representing a planetary body) to investigate an impactor/target mass ratio below 1\% \citep{carter2018collisional}. This represents a strong limitation to this approach as planetesimals likely contribute to the majority of impacts and collisional erosion may be driven by small impactors. Accordingly, the physics of smaller cratering impacts \citep{holsapple2007crater, housen2011ejecta, svetsov2011cratering}  needs to be taken into account to better assess the effects of crustal stripping on the chemical composition of the BSE.  \cite{allibert2021quantitative} have proposed a model taking into account such a contribution from small planetesimals to the crustal stripping. They however neglect the possible contribution from already fractionated embryos further impacting the proto-Earth: all embryos in their model are considered as chondritic in composition. That is a strong limiting assumption that is known to be wrong  in the case of substantial erosion previously experienced by these embryos. An improvement of this model is thus required to account for this effect. In addition, they have produced results focusing only on the Sm-Nd couple. However, all Earth RLEs may be fractionated in the BSE due to collisional erosion. They should thus be explored. The consequences of such fractionation, once estimated, should also be discussed with respect to terrestrial planets accretion scenarios. The implications this has on the nature of terrestrial planets parent bodies must as well be discussed. As such, our work proposes a significant improvement compared to previous studies:  
\begin{itemize}
    \item The impact-induced fractionation of different couples of RLEs, relative to one another, are tracked for about 100-200Myr for Earth analogs
    \item The loss (relative to primitive material) of individual RLEs is estimated as well
    \item The particular case of the popular Grand Tack scenario is studied in depth, in particular the influence of the date of the Moon-forming event
    \item The influence of small planetesimals is considered
    \item A coupling tracking all embryos together to account for their own fractionation and further implications in impacts is computed
    \item The influence of the partial melting rate in crustal formation history is also explored
\end{itemize}

\section{Methods}

 The collisional \label{sec:methods} history of every embryo is tracked with N-body numerical simulations \citep{jacobson2014lunar} of planetary accretion in the context of a Grand Tack scenario, that is among the most popular scenarios for planetary formation. The Grand Tack indeed invokes inward (followed by an outward) migration of Jupiter and Saturn that could notably explain the small size of Mars. The mass transfer during an impact is computed with an analytical model for each impact. A geochemical mass balance is additionally performed to re-compute the composition of the embryo after each impact. Integrating this procedure over entire collisional histories of growing embryos corresponding to Earth analogs allows a quantification of the final fractionation of various elements within the crust and mantle of each embryo (Bulk Silicate Planet - BSP). A more detailed description of the model is provided in appendices. It is based on the model presented in \citep[][]{allibert2021quantitative}, but adds the coupling between all embryos in the system (they are simultaneously tracked). This is a strong addition to the model since \cite{allibert2021quantitative} showed that small embryos may be strongly fractionated in RLEs. In this case, their further contribution to the proto-Earth accretion may cause further fractionation in Earth's primitive mantle. This effect has to be taken into account and properly quantified. \newline

\subsection{Model description}

Basic concepts on the modelling used in this study are described in \citep[][]{allibert2021quantitative}. A brief summary of the model, along with its new implementations are presented here. \newline 

The \label{sec:methods} collision files from N-body numerical simulations provide a record of every single impact occurring during a simulation with its associated parameters (velocity, angle, time, impactor to target mass ratio). Impacts can be of two types: either embryo-embryo (E-E) collisions or embryo-planetesimal (E-P) collisions. In the (E-P) case the test particle representing a planetesimal stands for a collection of a large number of planetesimals with the same total mass as the given test particle. Accordingly, our model replaces the test particle by a large number ($>$1000) of smaller planetesimals with a size distribution following a power law with $dn/dr \propto r^{-3.5}$ \citep{dohnanyi1969collisional, bottke2005fossilized} with same geometry and velocity at infinity as the test particle. 

The N-body simulations last for 100 Myr - 200 Myr and are assumed to start when the embryos are formed, before the onset of giant planets migration (i.e. 1-2 Myr after CAIs following \cite{walsh2011low}). The initial system contains a population of embryos and two distinct populations of planetesimals (internal planetesimals - within 3 AU and with a mass of $3.8\times10^{-4}M_\oplus$ - and external planetesimals, with a smaller mass of $1.2\times10^{-4}M_\oplus$) \citep[referenced as ``1to1'' simulations in][]{jacobson2014lunar}. Each simulation begins with approximately 100 embryos and 2000 planetesimals represented by test particles. The total mass of the system is equally distributed between the different planetesimals and embryos: total planetesimals mass is identical to the total embryos mass \citep{jacobson2014lunar}. However, the typical mass of an embryo at the beginning of the simulation (i.e. the typical mass of an embryo after the onset of the inward migration of Jupiter roughly 1Myr after CAIs formation) is a free parameter. Three different values are tested: 0.025$M_{\oplus}$, 0.05$M_{\oplus}$ or 0.08$M_{\oplus}$. Embryos are assumed to be differentiated, while planetesimals are assumed to be undifferentiated. Giant E-E impacts and  E-P impacts are treated in two distinct ways following:\newline
\begin{itemize}
    \item E-E: perfect merging is assumed for this type of impacts. The accreted material to the growing Earth embryo has the same composition than the impactor. The BSP is fully melted and crust and mantle compositions are re-equilibrated. The newly formed crust is re-enriched in incompatible elements according to their solid-liquid partition coefficients \citep{workman2005major}. The melting rate during crust-mantle differentiation is fixed to 2.6\% following the value proposed by \cite{o2008collisional}.  This value is estimated based on three constraints added to a 2-stage model for collision erosion: 1- differentiation into a crust and a mantle and 2- crustal stripping during a single impact. The three constraints are the following: 
    \begin{itemize}
        \item (i) the Earth Fe/Mg ratio must be fitted,
        \item (ii) the amount of $^{40}$Ar in the atmosphere (resulting from K decay) is taken to fix the loss limit in the most incompatible element to 0.5 and
        \item  (iii) the $^{142}$Nd/$^{144}Nd$ ratio measured in excess in terrestrial rocks compared to most chondrites must be explained fully by collisional erosion.
    \end{itemize} 
    Even if the third constraint is not favored anymore nowadays, we believe that the 2.6\% partial melting value reflects as a good starting point offering  access to the upper bound of the chemical fractionation produced by crustal stripping during impacts (cf (ii) notably). Higher partial melting rates are expected to be more realistic, especially during the early phases of planetary formation. However, this parameter is still under-constrained, such as the exact process leading to early crusts formation. Accordingly, we make the choice of providing estimates of the fractionating power of collisional erosion for two given end-members. Most of the results will be presented for the higher fractionating power end-member (2.5\% partial melting). However, some additional results are provided for the second end-member whose choice is justified hereafter. Depending on the ratio of the partitioning coefficients for two given elements, collisional erosion can have various effects on the fractionation for different partial melting rates. A higher partial melting rate will involve a smaller fractionation, for a given chemical element couple. It is even possible in certain cases to imagine no fractionation at all because of the high partial melting rate leading to low incompatible elements enrichment in the crust. However, that still depends on the respective partitioning coefficients of the chemical elements of focus. The higher the partitioning coefficients ratio, the higher the melting rate must be before collisional erosion have no more effect on the relative fractionation. As an example, for a ratio of 5 between the partitioning coefficients of two given elements, collisional erosion would have no effect if the partial melting rate is higher than $~$20\%. As mentioned, 2.6\% partial melting rate is a rather low value that is unlikely to have been neither constant nor this low during the entire 200 Myr tracked in this study. Notably, during the first millions years of accretion, the partial melting rate to form early planetary crusts are thought to be possibly up to 20\% based on estimates on Vesta, or between 5\% to 10\% for Ceres or the Moon \citep{yamaguchi1997metamorphic, ruzicka1997vesta}. The particular case of the growing Earth is still under-constrained. One could suspect that due to the existence of magma oceans, higher geotherms and potentially more H$_2$O in the embryos and proto-planets would induce a higher than 3\% melting rate. However, it seems unlikely that a 20\% partial melting rate (as suggested from Vesta) may have remained during the entire accretion history, especially during the late phases of accretion while the strong radiogenic contribution from Al$_{26}$ has disappeared and less collisions occur. The presence of more H$_2$O also strongly depends on the embryos considered and on their specific history (past accreted material, but also initial location in the disk). To propose a second end-member to the possible fractionation in refractory and lithophile elements, we thus explore the extreme case of a 20\% partial melting rate kept constant all along the accretional history. This provides a lower bound of the possible impact-induced fractionation on RLEs in the BSE. Note that the model does not take into account the core partitioning. 
    \item E-P: Here, the estimates of eroded and accreted masses are computed. These two masses depend on the impact parameters (i.e. velocity, angle, the masses and densities of the target and impactor) and are calculated using the parameterization from numerical simulations in \cite{svetsov2011cratering} in which the average (over angles) target losses is estimated as: 
    \begin{equation}
        \frac{m_{esc}}{m_{tar}} = 0.03 \left(\frac{V_{imp}}{v_{esc}}\right)^{1.65}\left(\frac{\rho_{tar}}{\rho_{imp}}\right)^{0.2}
    \end{equation}
    with $m_{esc}/m_{tar}$ the relative mass loss of target material, $V_{imp}/v_{esc}$ the impact to escape velocity ratio, $\rho_{tar}$ the target density at contact (here the crust) and $\rho_{imp}$ the impactor's density. For an ejected mass lower than crustal mass, the eroded material has a crustal composition. For larger ejected masses, the ejected material is made of both crust and mantle. The mass of ejected mantle is estimated as the difference between the total ejected mass and the entire crust mass. The accreted material is assumed to be chondritic in composition, that means that all accreted material as the initial composition of the growing embryos BSP. The absolute initial abundances in different elements will not influence the absolute loss of a given element since the results are normalized to their  initial contents. No new crust is formed. All the accreted material is thus deposited on the surface and the composition is diluted. N-body simulations are under-resolved: to increase the accuracy of the erosion process, each planetesimal in the N-body simulations is replaced by a population of smaller planetesimals of equivalent mass following a size-frequency distribution (see appendices). 
\end{itemize}

We define the parameter $\epsilon$ describing the chemical fractionation of the $M_1/M_2$ ratio ($M_1$ and $M_2$ being two given chemical elements) of the BSP at a given time $t$ with respect to its initial chondritic composition:
\begin{equation}
    \epsilon=\frac{\left(\frac{M_1}{M_2}\right)_{BSP}^{t}-\left(\frac{M_1}{M_2}\right)_{BSP}^{ini}}{\left(\frac{M_1}{M_2}\right)_{BSP}^{ini}}.
\end{equation}
We consider as relevant Earth analogs at the end of the N-body simulations the surviving embryos with a mass comprised in between $1/2M_{\oplus}$ and $1.2\times M_{\oplus}$ and final semi-major axis in between 0.5 AU and 2 AU. For each of them, $\epsilon$ is computed. The relative final to initial composition of a given element $M$ is also computed and denoted as $C_M^{BSP}/C_M^0$ for all Earth analogs and for the following elements: Rb, Ba, Th, U, K, Pb, Sr, Nd, Hf, Sm, Cs, Lu, Na, Si, Mn and Mg. 
    
\subsection{Selection of elements}
 The justification of the choice of elements chosen to apply the model (EROD) presented in section~\ref{sec:methods} is presented here. These chosen elements are the following:
\begin{itemize}
    \item (1) The pairs of refractory lithophile elements (RLEs) Sm-Nd, Lu-Hf and Th-U. Terrestrial planets including the Earth are assumed to have refractory lithophile element (RLE) ratios identical to those found in CI chondrites. This sets the geochemical paradigm that bulk differentiated planet compositions in refractory elements were established early (i.e. formation of chondritic parent bodies) and remained chondritic regardless of their accretion and differentiation histories. This chondritic paradigm is sustainable only if energetic collisions during the accretion history of terrestrial planets lead to small fractionation of RLEs. In this framework, Sm-Nd and Lu-Hf radiogenic isotopic system can provide precise constraints for defining an allowable space of conditions during Earth's accretion that would lead to the present estimates of $^{142}$Nd/$^{144}$Nd and $^{176}$Hf/$^{177}$Hf. U and Th are RLEs which are highly incompatible and their bulk Earth composition is accordingly particularly sensitive to the process of preferential stripping of the proto-crust. A substantial loss of these elements could be problematic for explaining the heat fluxes at the Earth surface that results {partly (about ~50\%)} from the radioactive decay of Th, U and K \citep{pollack1993heat}.
    \item (2) Moderately volatile elements such as Mn, Na, Rb,  Cs or K that are depleted with respect to chondritic compositions. The origin of these depletions can be attributed to incomplete condensation in the solar nebula and/or post nebular processes such as vaporization, differentiation, and collisional erosion. Thus, the potential effects of collisional erosion on the budget of moderately volatile elements needs to be estimated quantitatively to provide better constraints on the origin of volatile elements depletion. 
    \item Finally, (3) Mg and Si that are major rock forming elements with middle levels of incompatibility which do not strongly fractionate during crust-mantle differentiation. The observed superchondritic Mg/Si ratio is likely to reflect the presence of Si into the core \citep[e.g.][]{jagoutz1979abundances, fitoussi2009si}. Another alternative could be a 40\% of present-day Earth mass accreted from a first generation of refractory-rich planetesimals already enriched in Mg compared to Si \citep{morbidelli2020subsolar} but collisional erosion have also been suggested as a powerful mechanism for removing the missing Si from the BSE \citep{boujibar2015cosmochemical} and need to be physically quantified.\newline
 \end{itemize}
In the present work, we describe our approach for modelling quantitatively crustal erosion during accretion, present our results for the studied elements with variable geochemical properties and  discuss their implications for the composition of the  BSE.     

    \subsection{Mixing embryos compositions during merging events}
   
   Substantial preferential erosion of the crust for some embryos has been evidenced sometimes early in the collisional history of terrestrial planets. This could lead to an early chemical fractionation of the different growing embryos BSP in the system. Consequently, the RLEs composition of the accreted material for E-E impacts could be fractionated compared to chondritic values. This effect is taken into account in our model by implementing a coupling between the different embryos. The chemical evolutions of all embryos in the system are then simultaneously followed.

\section{Results}

In this section, we present the results of the chemical fractionation with respect to chondrite composition induced by collisional erosion expressed with $\epsilon$ for the following pairs of elements: Lu/Hf, Mn/Na, Sm/Nd, Rb/Sr, Th/U and Mg/Si. The results are summarized in table~\ref{tab:conclusion}. The average $\epsilon$ is estimated (a) for all Earth analogs and (b) only for Earth analogs that experienced a last giant impact after 50Myrs. 

\subsection{Evolution of the fractionation in one example simulation}

\begin{figure}
		\centerline{\includegraphics[width=1.1\linewidth]{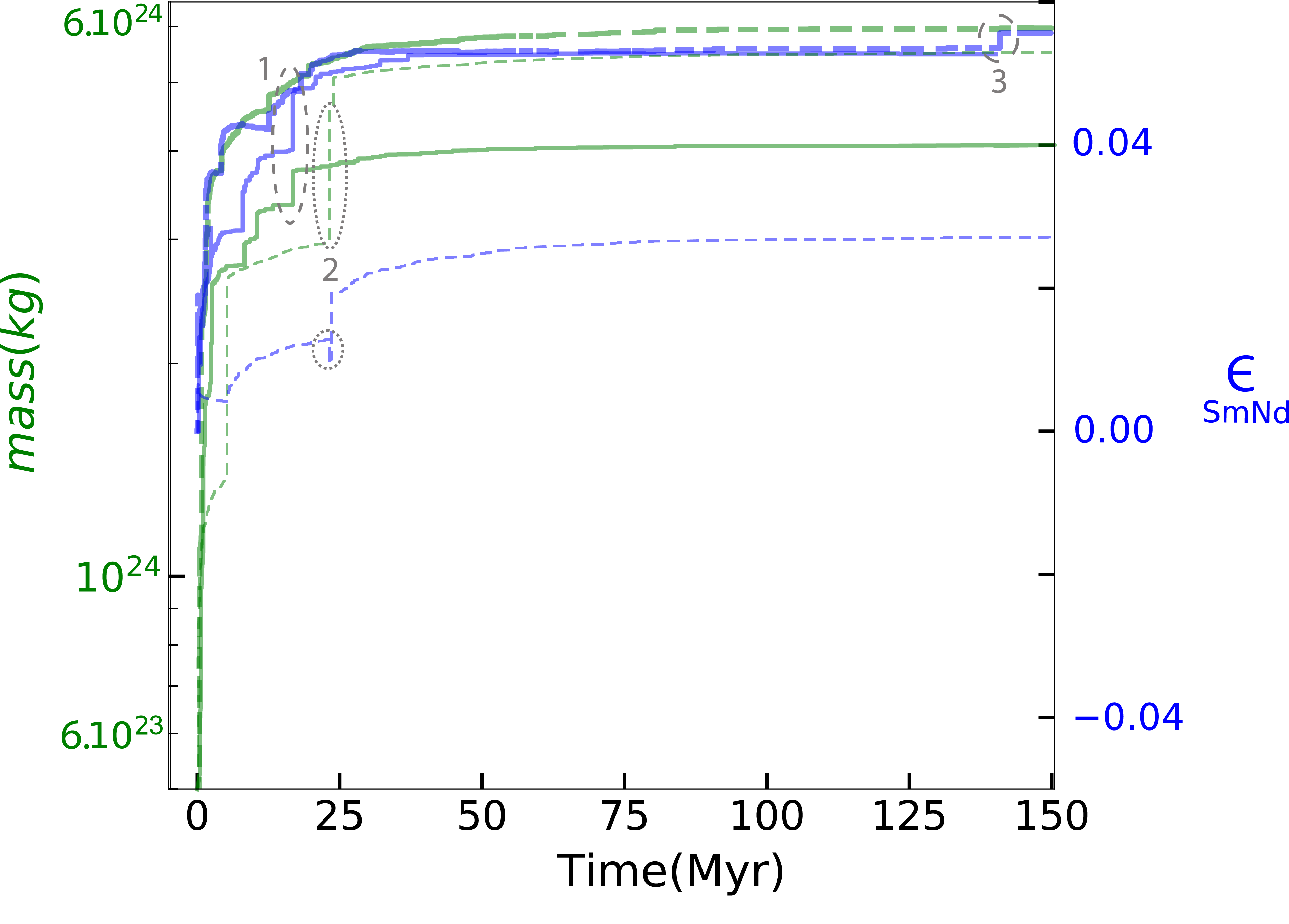}}
		\caption{Examples of both mass and $\epsilon$ (Sm/Nd) evolution with time for 3 embryos extracted from the N-body numerical simulations. The line style denotes the embryo identities whereas the line color denotes either the associated mass (in green) or the associated $\epsilon$ (in blue). The plain line, dashed line and pointed line are meant to denote a given embryo. (1) and (2) highlight two possible behavior in terms of fractionation for a giant embryo impact and (3) illustrates a more unusual case in which late fractionation occurs for  specific impact conditions (a high velocity for example, here more than 20km/s). 
		}
		\label{fig:evol}
	\end{figure}

The evolution with time of the chemical fractionation of the growing BSP composition is followed during the entire accretion history. $\epsilon$ is highly dependant on several parameters, notably the impact parameters as well as the chemical composition of the two bodies involved in the collision. We present here the cases of three different embryos evolving in different simulations to illustrate the typical evolution of the BSP composition with time as a response to their collisional histories. These three accreting histories are shown together in figure~\ref{fig:evol} which displays the fractionation of the Sm/Nd ratio as a function of the mass of the growing embryos. As Nd is more incompatible than Sm, preferential erosion of the crust with respect to the mantle increases the Sm/Nd of the BSP relatively to the initial chondritic ratio (and $\epsilon$, the fractionation value, increases consequently). This figure highlights the typical effect of a given impact whether a giant E-E or a E-P impact occurs. Overall, only a small fraction of the impacts have a significant influence on the fractionation which appears to be mainly driven by the impact velocities: the higher the velocity, the more material is typically ejected. However, successive impacts from planetesimals produce a significant fractionation, especially during the first million years of the simulations when the system is dynamically excited by the migration of the giant planets. Additionally, the less massive the embryo the easier the impactors erode the surface, leading to a higher fractionation in Sm and Nd of the BSP. However, after the first million years, the fractionation tends to either remain constant or to decrease. This is due to the amount of accreting planetesimals that have a chondritic composition. The crustal composition becomes more and more chondritic, leading to low fractionation even if some fraction of crust is still eroded. Regarding the role of E-E impacts, two main outcomes can be expected and they are highlighted in figure~\ref{fig:evol} by the numbers (1) and (2). (1) denotes the case of an impactor that has already experienced erosion and fractionation during its growth. In that case, the addition of very fractionated material to the BSP from the impactor induces a large increase in its fractionation. (2) refers to the case of an impact of an embryo with a chondritic composition. The direct result is a large increase of mass associated to a large decrease of $\epsilon$. Finally, (3) shows an other interesting behavior that can induce a significant fractionation even at late stages of accretion from a high velocity planetesimal-embryo impact (here higher than 20km/s). 
  \newline
 After following simultaneously the evolution of all embryos, the average final fractionation $\epsilon$ for embryos corresponding to relevant Earth analogs (i.e. relevant mass and distance to the sun) is computed for the selected set of elements. The most fractionated couple of elements is Rb/Sr for which $\epsilon=-$0.16$\pm$0.06 (all Earth analogs), followed by Lu/Hf with an average $\epsilon$ of 0.09$\pm$0.04. The less fractionated ratio is Th/U with $\epsilon=-$0.008$\pm$5.10$^{-4}$. This in an expected behavior according to the liquid/solid partitioning coefficient ratios of these couples (see table~\ref{tab:conclusion}). Note that a negative value corresponds to a preferential loss of the element on the numerator of the ratio contrarily to a positive $\epsilon$ value corresponding to a preferential loss of the element on the denominator. The table~\ref{tab:conclusion} also shows that a late last giant impact seems to induce a higher fractionation for all couple of elements. This behavior is further explored hereafter for all selected elements.

\subsection{Influence of the timing of the last Giant impact}

\begin{figure}
		\centerline{\includegraphics[width=1.1\linewidth]{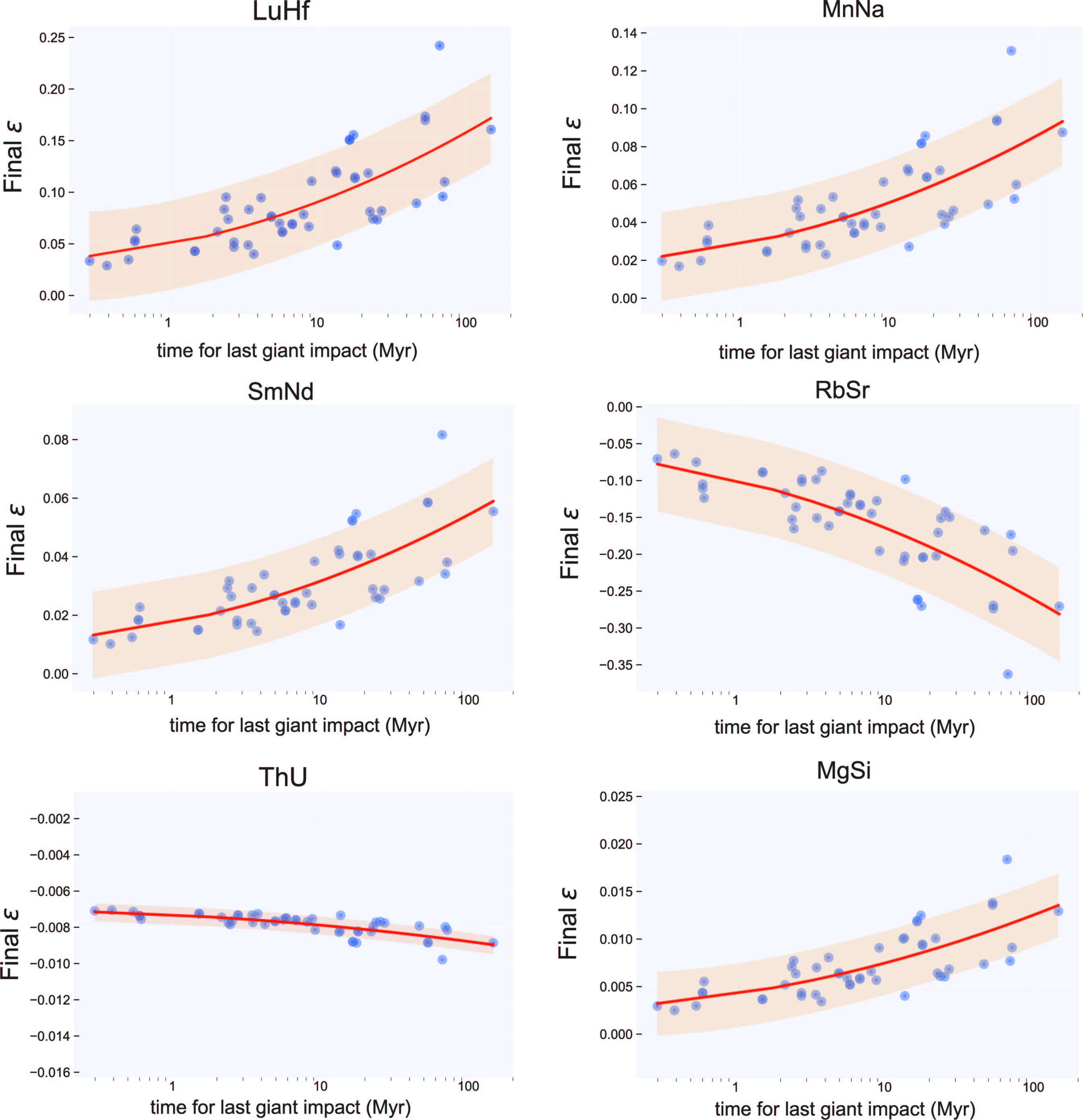}}
		\caption{Panel presenting the respective final $\epsilon$ fractionation for Earth analogs as a function of the timing of the last giant impact for the 6 different couples of chemical elements. The red shadowed region refers to the fitted (by least square) trend of the evolution of the fractionation as a function of the timing of the Moon forming impact. Light orange region refers to the 1$\sigma$ standard deviation of $\epsilon$ in our model. 
		}
		\label{fig:panel}
	\end{figure}

The final ratios of given couples of elements (Lu/Hf, Mn/Na, Sm/Nd, Rb/Sr, Th/U and Mg/Si) estimated with our modelling are showed in fig.~\ref{fig:panel} which reports $\epsilon$ as a function of the timing of the last giant impact for Earth analogs. There is a clear correlation between the fractionation value $\epsilon$ and the timing of the last giant impact experienced by an embryo (fig.~\ref{fig:panel}). For all ratios, the later the last giant-impact occurs, the higher $\epsilon$ is. The most incompatible elements are the most strongly depleted among the different ratios. Indeed, the influence of erosion on the fractionation is highly dependent on crustal composition and accretion history of the Earth. In the Grand Tack scenario, the embryos are growing fast and reach a mass comparable to their final mass in roughly 20Myr. After that, most of the accreted mass comes from the population of planetesimals (E-P impacts). During crust forming events (E-E impacts), the crust is highly enriched in incompatible elements, and deviates from chondritic composition. Conversely, in the case of E-P impacts, because no new crust is formed, the delivery of a large mass of chondritic components brings back the crustal composition closer to the  chondritic composition. In that case, even an efficient erosive evolution does not strongly influence the final BSP composition. The more the chondritic accreted mass is between two giant impacts, the harder it is to fractionate the BSP. Further fractionation can only occur after a melting event produces a new crust, re-enriched in incompatible elements. The model assumes a constant partial melting rate for newly formed crust. Accordingly, the mass of crust increases as fast as the mass of the planet. It means that the mass of accreted material to the crustal mass ratio decreases with time, decreasing accordingly the significance of the accreted material composition. This makes it more difficult to lower the fractionation due to accretion of chondritic planetesimals as time increases. In this context, a late giant impact implies a higher efficiency to fractionate the BSP because of a late re-equilibration of the entire BSP composition leading to a re-enrichment of the crust in RLEs. 

\section{Discussion}

\subsection{Refractory and Lithophile Elements ratios}

\begin{figure}
		\centerline{\includegraphics[width=.6\linewidth]{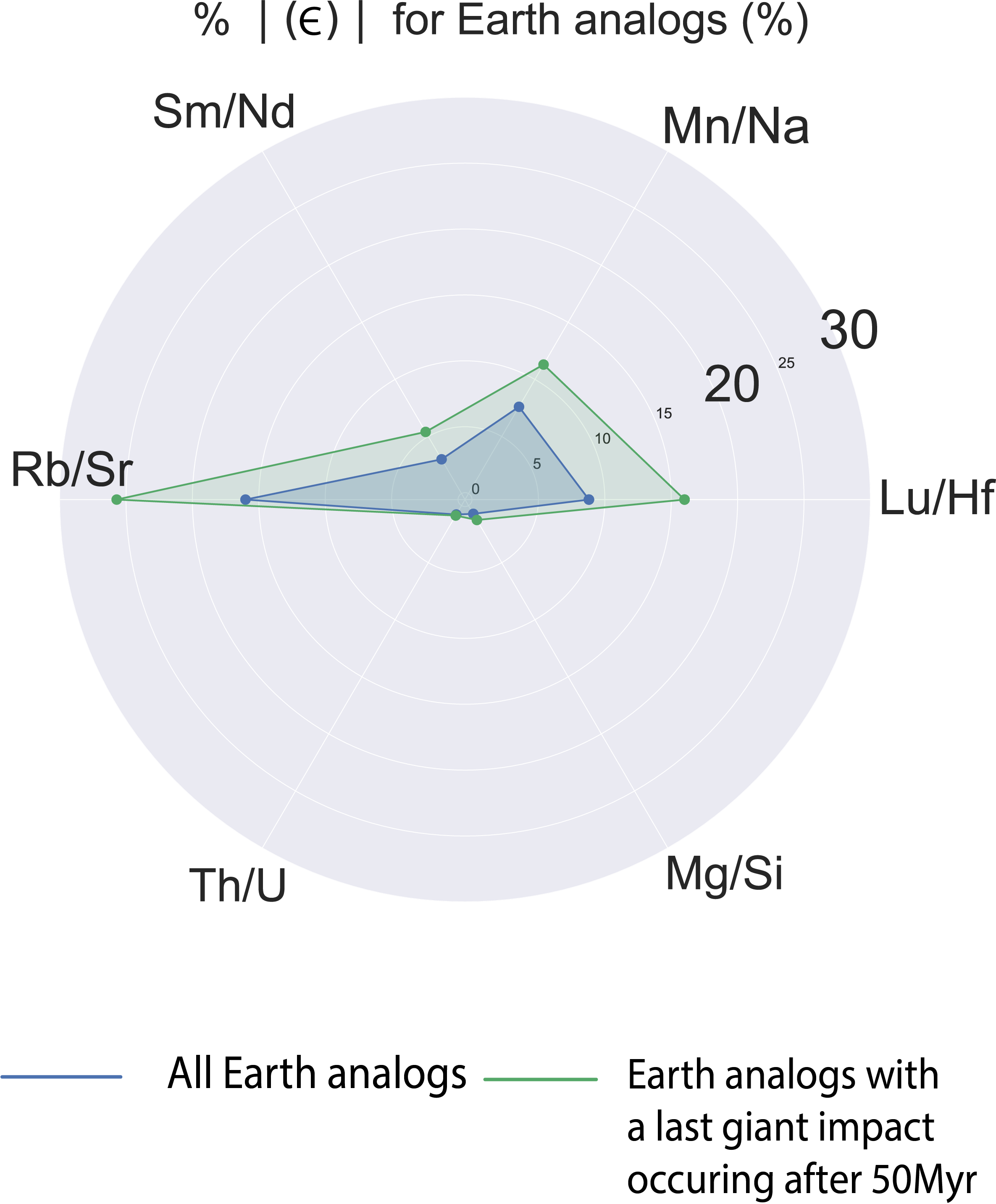}}
		\caption{Spider chart illustrating the respective $\epsilon$ values (in \%) for the relative fractionation between two elements for these 6 couples of elements: Lu/Hf, Mn/Na, Sm/Nd, Rb/Sr, Th/U and Mg/Si. These ratios are averaged from final ratios of relevant Earth analogs in N-body numerical simulations. The blue line corresponds to the average values obtained from all Earth analogs at the end of the simulations. The green line corresponds to the average values obtained only for Earth analogs that underwent a late ($>$50Myr) last giant impact. This later case is considered to be the most likely case based on geochemical and cosmochemical constraints\citep{chyba1991terrestrial, walker2009highly, touboul2007late, pepin2006xenon, norman2003chronology, jacobson2014highly}.
		}
		\label{fig:Ratios_RLEs}
	\end{figure}

The Th/U ratio remains unfractionated over the entire Earth collisional history (fig.~\ref{fig:Ratios_RLEs}); in agreement with the Th–U–Pb isotope systematics in terrestrial rocks \citep[e.g.][]{sun1989chemical, wang2011geochemistry}. This is due to the very low difference between the degrees of incompatibility of Thorium ($D_{Th}=0.001$) and Uranium ($D_{U}=0.0011$).\newline

Sm/Nd ratio is fractionated during accretion. This has already been highlighted by \cite{allibert2021quantitative} who showed that almost 5\% fractionation in the Sm/Nd ratio is possible in the case of a Grand Tack and a late last giant impact. However, they neglect the fact that impactors may already have been fractionated, which needs to be taken into account. For a late last giant impact, the fractionation $\epsilon$ value reaches 0.054 ($\pm 0.016$) in the current model, close to the 6\% fractionation invoked to account for the super-chondritic Sm/Nd ratio. This is a higher estimate as proposed by \cite{allibert2021quantitative}, but from only 0.05\%, showing the relatively low contribution to the fractionation from the differentiated impactors. \newline 
The initially proposed superchondritic Sm/Nd ratio could, a priori, be fully explained by collisional erosion. However, for a 6\% increase in the Sm/Nd ratio to account for the possibly superchondritic $^{142}$Nd$/^{144}$Nd, this fractionation should have happened within the first 30 Myr of planetary formation \citep{boyet2005142nd}, which is inconsistent with the only scenario capable of producing such a large fractionation: the one that invokes a late ($>$50 Myr) last giant impact on Earth. Additionally, as preserved nucleosynthetic anomalies \citep{burkhardt2016nucleosynthetic, bouvier2016primitive} within the disc represent a strong alternative for explaining the 20-ppm excess in $^{142}$Nd compared to $^{144}$Nd (instead of an excess of Sm compared to Nd), the combination of a Grand Tack and a late moon forming impact may be incompatible with the current understanding on the $^{142}$Nd$/^{144}$Nd data in Earth samples because of producing a too large, and too late, fractionation in Sm and Nd. However, considering the new light shed by \citep{frossard2022earth}, a fractionation in the Sm/Nd ratio of about 2.4\% by collisional erosion could have happened. That value is close to the 3\% estimated here in the case of Earth analogs evolving in a Grand-Tack scenario but without a late Moon-forming event. The current ranges of Moon-forming event dates do not extend under 30Myr \citep[e.g.][]{barboni2017early, jacobson2014highly, yin2014records, bottke2015dating}. This suggests that under the assumptions made here, in the context of a Grand Tack scenario, the Sm/Nd ratio could help constraining the date of the Moon-forming impact, placing it here between 30Myr and 50Myr. It is however important to keep in mind that the crust formation processes are still too under-constrained at the time to be able to make such a strong point. It only argues for the power of this tool in the future into shedding light on events such as the Moon-forming event and accretion scenarios in general.  \newline

Lu-Hf isotope system shares similarities with the Sm-Nd system. The Lu-Hf ratio of the BSE is estimated after Sm-Nd systematics of chondrites and terrestrial rocks \citep{jackson2013major}. For a maximum 5\% to 8\% fractionation of the Sm/Nd ratio, a 12\% fractionation is predicted for the Lu/Hf ratio of the BSE \citep{bouvier2008lu}. Such value can be obtained in the case of a Grand Tack dynamic model of planetary accretion combined with a late moon forming impact ($\epsilon$(Lu/Hf) $= 0.16 \pm 0.05$). Therefore, collisional erosion could produce superchondritic ratios of both Sm/Nd and Lu/Hf in the BSE in this specific case. However, the Grand Tack combined with a late Moon forming impact would produce the most extreme possible, if not too extreme, fractionation acceptable for both Sm/Nd and Lu/Hf ratios and more generally would produce substantial chemical fractionation among other refractory lithophile elements (RLEs). This could impose restrictions on the validity of combining these two popular hypotheses for the understanding of Earth's formation that conveniently explain some chemical observables (e.g. for Hf/W, Mg-suite crustal rocks, highly siderophile elements, I-Xe) \citep[e.g.][]{chyba1991terrestrial,pepin2006xenon, touboul2007late, walker2009highly}, the dating of lunar rocks and estimates of its formation \citep[e.g.][]{norman2003chronology,  borg2011chronological} or the outputs of numerical simulations \citep{jacobson2014highly,jacobson2014lunar}. On the other hand, the occurrence of a Grand Tack scenario with a late moon forming impact (after 50 Myr) would inevitably produce notably non-chondritic BSE compositions in refractory lithophile elements. Quantifying the effects of impact induced erosion on other chemical elements with varying geochemical properties is critical to set the limits of elemental loss during planetary accretion and provide estimates on the BSE non-chondritic composition resulting from collisional erosion eventually. \newline 

\subsection{The BSP composition resulting from collisional erosion}

\begin{figure}
		\centerline{\includegraphics[width=1.\linewidth]{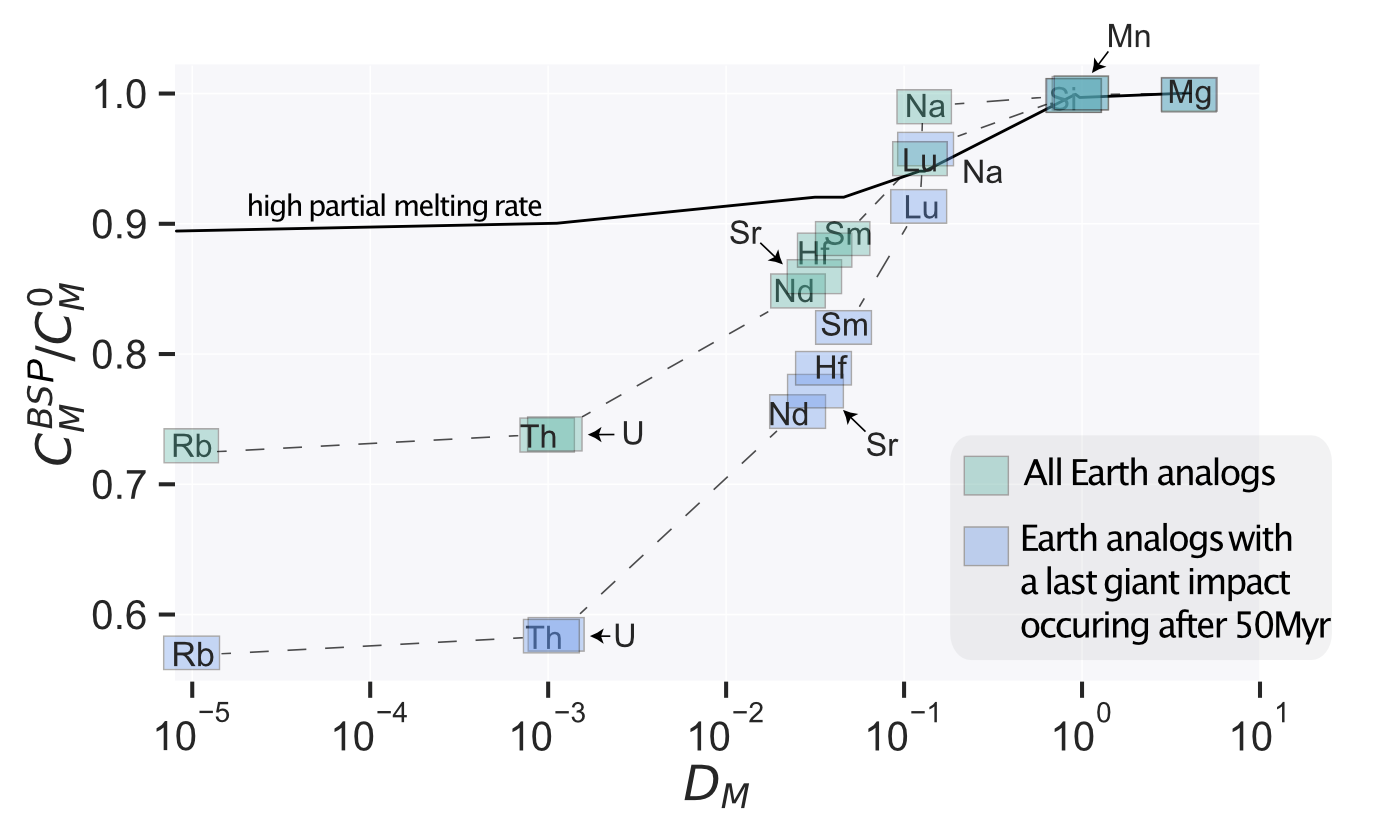}}
		\caption{Effects of crustal stripping on the BSP final composition with respect to initial chondritic composition. The elemental depletion of an element M $C_M^{BSP}/C_M^0$ ratio is displayed as a function of the decreasing degree of incompatibility (liquid-solid partition coefficients are taken after \citep{workman2005major}. $C_M^{BSP}$ corresponds to the average calculated final concentration of M in the BSP after collisional  histories of proto-Earth are completed. $C_M^0$ refers to the initial chondritic concentration of the embryo at the beginning of the last stages of accretion (roughly 2-3Myr after CAIs). An additional case is added: the case of a constant 20\% partial melting rate over the entire collisional history (black plain line). 
		}
		\label{fig:BSE_compo}
\end{figure}

From the modelling of collisional erosion during Earth accretion histories, we infer the final BSP concentrations in Rb, Th, U, Sr, Nd, Hf, Sm, Lu, Na, Si, Mn and Mg relatively to their initial chondritic concentrations as a function of their degree of incompatibility (fig.~\ref{fig:BSE_compo}). As expected, the depletion of considered elements compared to initial values is observed for an increasing degree of incompatibility (decreasing $D_M^{sol-liq}$; from Lu to Rb). Major elements (Mg, Si) are not affected by collisional erosion. Moderately incompatible  elements are either not affected (Mn) or poorly affected (Na). We find that the occurrence of a late giant impact leads to higher depletion of incompatible elements. Additionally, the difference of depletion in between all models of Earth analogs and models with a late giant impact increases with the increasing degree of incompatibility of elements. Most incompatible elements such as Rb reach a final depletion of $C_M^{BSP}/C_M^0 = 0.57$ in the case of a late giant impact, which corresponds to the highest loss obtained from our modelling (fig.~\ref{fig:BSE_compo}). As a comparison, the two stages model of collisional erosion proposed by \cite{o2008collisional} (i.e. (1) formation of a crust enriched in incompatible elements; (2) erosion of 54\% of this crust during a single impact) imposes a limit of $C_M^{BSP}/C_M^0$= 0.5 for the most incompatible elements. Therefore, the maximum elemental loss from an entire collisional accretion history might be less important than the maximal loss assumed in \cite{o2008collisional} from a single erosive impact event that would strip away more than 50\% of crust. This is even more true when considering the case of a high partial melting rate (black line in fig~\ref{fig:BSE_compo}) for which less fractionation is observed (only about 10\% for the most incompatible element, Rb). For elements that have a moderate degree of incompatibility, such as Lu or Na, there is only very low difference in the fractionation whatever the degree of melting. Major elements with a low incompatibility like Si, Mg, are also not affected by collisional erosion in this case. This illustrates how important the partial melting rate is, as it decreases significantly the fractionation, even for the most incompatible elements such as U and Th.\newline

\begin{figure}
		\centerline{\includegraphics[width=1.\linewidth]{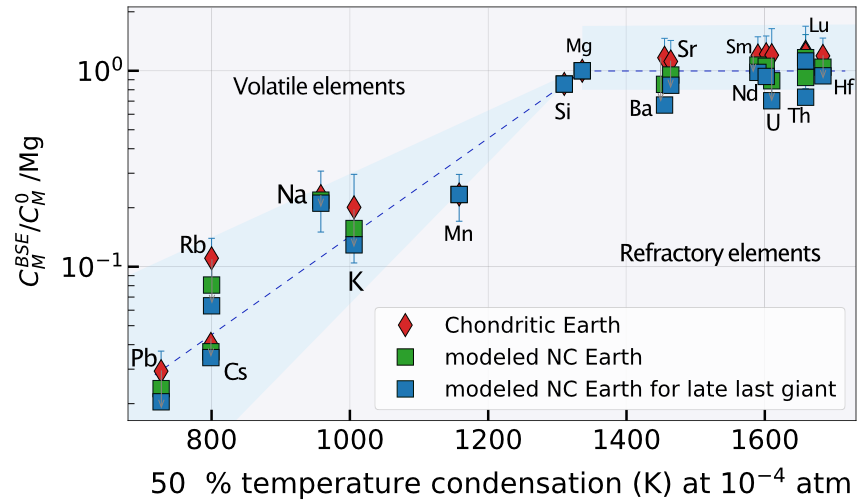}}
		\caption{Abundances of elements (Pb, Cs, Rb, Na, K, Mn, Si, Mg, Ba, Sr, Sm, Nd, U, Th, Lu and Hf) in the bulk silicate Earth normalized to CI chondrites and Mg as a function of their 50\% condensation temperature. Red square symbols lie for current estimates based on a chondritic Earth model along with their 1$\sigma$ uncertainties \citep{palme2003evidence, barrat2012geochemistry}. Green and blue squares symbols show the corrected values for BSE abundances when the effect of collisional erosion modeled in our work are taken into account. The green symbols show the average abundances obtained for all the Earth analogs issued from N-body numerical simulations. The blue squares correspond specifically to the Earth analogs recording a last giant impact after 50 Myr. These results are produced for a constant partial melting rate of 2.6\%.
		}
		\label{fig:Tcond}
\end{figure}

Using these estimates of elemental loss, we have computed the BSE final abundances (normalized to CI and to Mg) corrected for the effect of collisional erosion. The results are shown in figure~\ref{fig:Tcond} for the studied set of chemical elements as a function of their 50\% condensation temperature. The chondritic BSE estimates are presented in blue squares \citep{palme2003evidence, barrat2012geochemistry} and are notably based on the assumption that refractory lithophile elements are not depleted with respect to chondritic compositions. The outcomes from our modelling are displayed for two distinct cases: (1) taking into account the entire set of Earth analogs surviving the N-body numerical simulations (green squares); (2) taking into account only Earth analogs recording a last giant impact after 50 Myr (red squares). These values represent the expected changes on BSE composition due to collisional erosion. Such a correction is of interest considering the fact that nowadays BSE composition estimates are mainly built on the assumption that RLEs in the BSE are in chondritic relative proportions by nature since they are not affected by devolatilization nor by core formation. This assumption may not be true in the case of an efficient collisional erosion. In particular, the estimates in the most incompatible elements can be very sensitive to this because the correlation with the major elements (such as Mg) composition in the most fertile peridotites can not be used to model their abundances. The corrected abundances values are calculated following:
\begin{equation}
    \left[\frac{X_{BSE}}{X_{CI}}-\left(1-\frac{X_{model}}{X_0}\right)\times\frac{X_{BSE}}{X_{CI}}\right] \times \frac{Mg_{CI}}{Mg_{BSE}}, 
\end{equation}
where $X_{BSE}/X_{CI}$ the abundance of X in the chondritic Earth normalized to CI composition estimated with CI and BSE compositions given in \citep{palme2003evidence, barrat2012geochemistry}, $X_{model}/X_0$ the average X abundance of the BSP of Earth analogs computed from N-body simulations normalized to their initial composition and $Mg_{CI}$ and $Mg_{BSE}$ being respectively the abundances of Mg in the CI chondrites and in the BSE chondritic model. For moderately volatile elements considered in this work (i.e. Pb, Cs, Rb, Na, K and Mn), the modeled values corrected from the effect of collisional erosion lie within errors (within 1$\sigma$ to 2$\sigma$) with geochemical estimates of BSE based on the chondritic assumption. Large uncertainties on both chondritic and BSE abundances for volatile elements \citep{palme2003evidence} make this statement even valid for the most incompatible elements such as Rb for which depletion induced by erosion can reach up to roughly 40\% in the case of low partial melting rate and a late last giant impact (fig.
~\ref{fig:BSE_compo}). Accordingly, integrated effects of collisional erosion to the current estimates of volatile elements show that this process seems unlikely to significantly contribute to the observed depletion of these elements on Earth with respect to their chondritic abundances. \newline

Incomplete condensation in the solar nebula \citep[e.g.][]{wasson1974fractionation} or partial vaporization from a molten magma ocean \citep[e.g.][]{hin2017magnesium, norris2017earth} should be regarded as the dominant processes for establishing the budget of volatile elements on Earth. Refractory lithophile elements (RLEs) are more notably affected by the process of collisional erosion as their predicted abundances can fall off the ranges of estimates based on the assumption of chondritic RLE ratios, notably for most incompatible RLEs such as U, Th and Ba. Taking into account the effect of collisional erosion on the BSE composition lead to (RLE/Mg)$_N$ (the elemental ratio in the BSE normalized ({\it noted N}) to Chondrites) that lies systematically below 1 (fig.
~\ref{fig:Tcond}) when all Earth analogs are considered. Such values fall below the 1 to 1.5 range proposed by most BSE geochemical models \citep[see tables in][]{palme2003cosmochemical, palme2014solar} but are compatible with the statistical approach developed by \cite{lyubetskaya2007chemical} that is meant to take into account how scatters in peridotite data affect the geochemical BSE composition estimates. For instance,     
  \cite{lyubetskaya2007chemical} finds 17.3$\pm$3ppb U content in the BSE  while 21.8ppb of U is given in \cite{palme2003evidence}. This leads within errors to a Mg and CI normalized value (U/Mg)$_N$ of 0.79 well below the 0.89 (U/Mg)$_N$ obtained from the present collisional erosion model. However, the (RLE/Mg)$_N$ values fall systematically below the lowest estimates from \citep{lyubetskaya2007chemical} for the most incompatible elements U, Th, and Ba if only simulations with a late moon forming impact are considered (e.g. (U/Mg)$_N$= 0.703 and (Ba/Mg)$_N$= 0.669). However, it is worth noting the fact that BSE estimates from \cite{lyubetskaya2007chemical} are not only different from any other study but also estimated with the same peridotite data base as \cite{mcdonough1995composition} which raises concern. This discrepancy may arise from the method used by \cite{lyubetskaya2007chemical} that introduces some bias in the resulting BSE compositions \citep{palme2003cosmochemical}. Accordingly, and regardless of the importance given to the data set from \cite{lyubetskaya2007chemical} in the interpretation, Earth formation scenarios including a last giant impact after 50 Myr could produce an impact induced chemical fractionation of some incompatible RLEs lying above any geochemical models of BSE composition. \newline 
  
   In this geodynamical context, the higher efficiency of collisional erosion could question the estimates of some highly incompatible RLEs based on the chondritic assumption and sampling of terrestrial rocks; or in contrary provide insights for deciphering the debated age of the lunar impact. For instance, Thorium (Th), Uranium (U) and Potassium (K) are radioactive RLEs and extensive loss of these elements during accretion would be at odds with the measured heat flux at the surface of the Earth. The heat fluxes at the Earth surface are essentially the result of two processes: (i) secular cooling, and (ii) radioactive decay of long-lived isotopes, mostly Th, U and K. However, the relative contribution of these two processes is still under-constrained and the estimates of the heat produced by radioactive decay can vary from up to a factor 20. As a result, the absolute compositions of Th, U and K in the BSE are poorly constrained and range roughly in between 10-25 ppb for U and 40-95 ppb for Th using either constraints from fluxes of geoneutrinos \citep{araki2005experimental, vsramek2013geophysical}, $^4He$ fluxes through the ocean floor \citep{o1983heat, o2008collisional}, or geochemical models \citep{palme2003evidence}. Taking the upper bound values from these discussed estimates for U and Th abundances would produce at max a 24 TW heat flux at the surface of the Earth \citep{dye2012geoneutrinos, vsramek2013geophysical}, much lower than the heat flux effectively measured ranging from 43 TW to 49 TW \citep{pollack1993heat, jaupart2007temperatures} and lower than the 33 TW predicted by geodynamical estimates \citep{vsramek2013geophysical} as to be the result of radioactive decay only. In the present work, crustal stripping in the context of both Grand Tack and late Moon forming impact could remove  about 40\% to 50\% of the heat producing elements U and Th with respect to chondritic composition, possibly increasing the discrepancy in between the heat flux and predicted abundances of these elements in the BSE. The modeled final U and Th compositions of $13 \pm 2$ ppb and $47.5 \pm 7.1$ ppb respectively would fit marginally with the lowest geochemical estimates for these elements. Considering all Earth analogs, the average loss of both Th and U is lower, comprised between 20\% and 30\%, providing better match with estimates of  U an Th contents in the BSE. In any case, we show that collisional erosion can be an efficient process for removing significantly the most incompatible RLEs (e.g. U, Th, Ba) and can produce non chondritic compositions of the BSE eventually (Fig.
 ~\ref{fig:Ratios_RLEs} (ratios) and~\ref{fig:BSE_compo} (absolute abundances)). As enlighten by the black plain line in~\ref{fig:BSE_compo}, this is the case at least for low partial melting rates, but chemical fractionation by collisional erosion may be more limited for higher degrees of melting. It also depends on the initial composition of the building blocks of Earth. If Earth building blocks were with concentrations in U under 20 ppb, then collisional erosion would be too efficient for allowing Earth to keep a sufficient amount of heat producing elements. In that case, a less erosive process of accretion than a Grand Tack with a late Moon forming event may be expected instead, as suggested by the estimations of U and Ba contents normalized to CI and Mg.  \newline
 If the most erosive case of a Grand Tack combined to a late moon forming impact appears not to be favored considering BSE estimates invoked above, there could however be other explanations for a low loss of the most incompatible elements. 
The partial melting rate is a critical parameter for the model. Accordingly, an alternative way for producing a low fractionation in the RLEs budget of the BSE could be to form crusts with a large amount of melt, leading to thicker crusts  with more diluted RLEs. \newline
Within one of the viable scenario (here considering no late last giant impact), can we deduce some other interesting features? 

    \subsubsection{Mg-Si as an indicator of loss for major elements}
         A superchondritic Mg/Si ratio has been reported for the BSE when compared to any type of chondritic material \citep[e.g.][]{jagoutz1979abundances, allegre1995chemical, fitoussi2009si}. This discrepancy is even more emphasized when enstatite chondrites are considered as the Earth's building blocks \citep{javoy1995integral, javoy2010chemical, boujibar2015cosmochemical}. Different hypotheses have been proposed to explain these non-chondritic Mg/Si ratio estimates such as the incorporation of Si into the core during differentiation, the loss of Si by volatilization during accretion, or an heterogeneous mantle with a Si-enriched lower mantle \citep{ringwood1958constitution, allegre1995chemical,fitoussi2009si,  javoy2010chemical}. Crustal stripping from impacts could also modify significantly the Mg/Si of the BSE as Mg and Si have different degrees of incompatibility as proposed in \citep{boujibar2015cosmochemical}.\newline
         Here we report a low (about 1\%) close to negligible effect of collisional erosion on the final Mg/Si ratio of the BSE. This is mostly due to their relative low degree of incompatibility which hamper a significant fractionation of these elements in crust forming liquids \citep{o2008collisional}. In this particular case, the amount of crustal preferential loss is not large enough to affect the Mg/Si of the BSE. This promotes the presence of Si into the Earth's core to account for the superchondritic Mg/Si of the BSE without any change to its predicted abundance from geochemical models \citep[e.g.][]{allegre1995chemical, javoy2010chemical}. As a result, our model which physically addresses the effects of accretion and erosion on the Mg/Si content of the BSE over a full accretion history leads to a different conclusion than the one proposed by \cite{boujibar2015cosmochemical} which modeled crustal erosion as a single stage impact event.
    
        \subsubsection{Effects of collisional erosion on the depletion of moderately volatile elements}
        
        The origin and timing of the depletion of volatile elements on Earth (see figure~\ref{fig:Tcond}) remain unresolved and play a key role for constraining the process of Earth's accretion and the nature of its building blocks \citep[e.g.][]{albarede2009volatile}. \newline
        The depletion of moderately volatile elements (MVEs) can be attributed to incomplete condensation in the solar nebula as well as post nebular processes including evaporation during the melting of proto-planets or storage into the core as most of the MVEs can also be siderophile under the conditions of core-mantle equilibration. In this picture, the role of collisional erosion on the observed depletion of volatile elements has been poorly constrained. For instance, Manganese (Mn) and Sodium (Na) are both moderately volatile elements with similar temperature of condensation at conditions of the solar nebula (low fO$_2$ and low pressure). As Na becomes more volatile than Mn under oxidizing conditions, the superchondritic Mn/Na ratio evidenced in some terrestrial bodies such as Mars has been used as evidence for volatile depletion from evaporation processes during planetary growth \citep{o2008collisional, campbell2012evidence, siebert2018chondritic}. As Na is more incompatible than Mn, collisional erosion can provide an alternative explanation for the superchondritic Mn/Na of differentiated planetary bodies. Moreover, Earth is the only differentiated body with a chondritic Mn/Na and estimating the effect of collisional erosion on these elements could help deciphering in between the effects that control the depletion of moderately volatile elements on Earth \citep{lodders1998planetary, o2008collisional, barrat2012geochemistry, siebert2018chondritic}. Note that the Mn/Na ratios of all classes of chondritic meteorites are nearly constant, almost within analytical error at the solar value of 0.39 $\pm$ 0.02, as derived from CI meteorites \citep{palme2014solar}. From our modelling, we infer a maximum relative loss of Na compared to Mn of $\sim 11\%$ for collisional erosion processes if the Moon formed after 50 Myr. This argues for a limited effect of post nebular processes (i.e evaporation and collisional erosion) on the fractionation of Mn and Na for the BSE in fair agreement with the observed chondritic Mn/Na.
    
        \subsubsection{The example of Rb/Sr as a tracer of the origin of Earth's volatile elements}
        Rb and Sr are both lithophile elements with partition coefficients between liquid and solid that are different by several orders of magnitude (i.e. Rb is a highly incompatible element). In addition, they have different volatility degrees (Rb is volatile while Sr is refractory). Therefore, the Rb/Sr ratio of the BSE is lower than the chondritic value \citep{halliday2001search} by an order of magnitude with a Rb/Sr of $0.030 \pm 0.006$ for the BSE while the chondritic value is estimated to be $0.32 \pm 0.10$ \citep{mcdonough1995composition}. Rb is one of the most incompatible elements among MVEs and can accordingly provide a strong insight on the maximum effect of collisional erosion on the depletion of MVEs. Such effect, if significant, should be taken into account when estimating the general depletion trend of volatile elements of the BSE. Here, we show that a preferential loss of Rb compared to Sr of roughly 25\% can be achieved by crustal stripping during Earth accretion over 200 Myr. This represents roughly 30\% of the total loss of Rb compared to chondrites inferred from measurements in terrestrial rocks and chondritic meteorites. We show that even for such a highly incompatible element, the origin of its depletion is mainly due to incomplete condensation or volatilization processes. As the abundances of volatile elements are poorly constrained in both the BSE and chondrites, the results show that estimates corrected from the effects of collisional erosion would only marginally affect the general depletion trend of volatile elements in the BSE. \newline

 The BSE abundances in volatile elements do not provide a strong constraint on the collisional erosion viability neither on dynamical scenarios favored. However, the incompatible RLEs such as U, Ba, Sm and Nd suggest that a low erosive process may be favored, at least in the framework of the current model and its assumptions.
 
 \subsection{Implications for the dynamics of terrestrial planet formation}
 
 The growth of terrestrial planets in such a low erosive environment requires the setting of a challenging dynamical context for terrestrial planet formation. A system containing both embryos and planetesimals requires, to be non-erosive, that their random velocities are low compared to the escape velocity of planetary embryos and/or that the collision rate is low. This can be achieved if the mass contained in the growing region of terrestrial planets is mainly distributed into large planetary embryos before the onset of dynamical excitation due to giant planets migration. This implies the presence of a low fraction of planetesimals with respect to planetary embryos early in the history of solar system formation (less than a few Myr) or a delayed instability of giant planets. In such a case, erosion is likely to be reduced despite the dynamical excitation induced by giant planets migration. Accordingly, the rapid formation of terrestrial planets via pebble accretion appears to be a suitable process \citep{lambrechts2014forming, johansen2015growth}. However, some N-body numerical simulations \citep{izidoro2016asteroid} suggest that a late instability would not likely reproduce the dynamical structure of the asteroid belt, as it would lead to an under-excited asteroid belt compared to current observations. \cite{kaib2016fragility} also disfavored the possibility of a late giant planet instability due to the excitation of terrestrial planets orbits. This last point is still a matter of debate since \cite{deienno2018excitation} revised this last result by showing that the asteroid belt could be excited by giant planets instability. A conundrum appears here: on one hand low erosion is required to explain the observed abundances of incompatible elements in the BSE (notably of the radioactive elements U and Th), while on the other hand a mechanism that would induce an efficient excitation is required to explain the structure of the asteroid belt \citep[see][for a review on the asteroid belt]{raymond2020origin}. A last possibility is that erosion could be over-estimated in the present model due to an oversimple description of the velocity distribution of fragments or due to the low partial melting rate chosen. This should be the purpose of future investigations.

\section{Conclusion}

This work presents a model ({\it EROD}) that quantifies the effect of collisional erosion on the abundances of elements in the bulk silicate Earth. It couples (1) semi-analytical simple laws for eroded and accreted masses during an impact event \citep{housen2011ejecta, svetsov2011cratering} and  (2) N-body numerical simulations of accretion in the context of Grand Tack \citep{jacobson2014lunar}. Resulting compositions are compared to the initial chondritic compositions for a set of elements with different geochemical properties: Sm, Nd, Lu, Hf, Th, U, Mn, Na, Rb, Sr, Mg, Ba, K, Cs, Pb and Si. The effect of crustal stripping during collisions on the chemical composition of the BSPs is strongly correlated to the partitioning coefficients of elements between liquid and solid silicates. A maximum loss of roughly 40\% is expected for the most incompatible elements (i.e. Rb, U, Th), if the Moon forming impact occurred late (i.e. 50Myr after CAIs) in the context of a Grand Tack accretion history. For major elements with a low degree of incompatibility (e.g. Si and Mg), the effect of collisional erosion is negligible. In between these extreme values, we provide a trend of depletion according to the degree of incompatibility of elements usable for correcting the present-day BSE composition models (for the most incompatible elements).

Additionally, we show that the BSE superchondritic Mg/Si ratio is unlikely to be due to crustal stripping. It emphasizes the need of an other fractionation process, possibly the incorporation of Si within the Earth core during differentiation, or nebular processes. The Mn/Na BSE ratio might be affected up to 11\% change due to collisional erosion and preferential loss of Na. This is a small but significant amount of change that strengthens the hypothesis of Mn partitioning within the core during its formation \citep{siebert2018chondritic}. This study outcomes concerning the Rb/Sr ratio show a 25\% change in the relative concentrations of Rb and Sr compared to chondrites that is due to accretion processes, including collisional erosion. That represents only a maximum of 30\% of the fractionation measured between the BSE and the chondrites, implying that the remaining loss might be due to devolatilization. This model may allow for the quantification of devolatilization processes on elements and improve the knowledge of the volatile elements depletion pattern of MVEs compared to chondritic compositions.

Finally, we find that in the context of a Grand Tack and a late Moon forming impact, the superchondritic $^{142}$Nd/$^{144}$Nd ratio could be explained by crustal stripping only. In that case, the Lu/Hf BSE ratio should be superchondritic by 16\%. However, in that later case, the loss in U and Ba (up to 40\%) is too large compared to expectations from BSE/CI/Mg estimates, excluding this scenario as a viable accretion process. This suggests a relatively low fractionation between Sm and Nd (about 3\% max), in agreement with the evidences of nucleosynthetic anomalies \citep{burkhardt2016nucleosynthetic, bouvier2016primitive, boyet2018enstatite, frossard2022earth}. These later estimates suggest that either the Earth has grown up into a low erosive environment possibly excluding the existence of a Grand Tack and a late Moon forming impact together. Another possibility could be to have formed the crust with a high partial melting rate at early times during accretion. In this regard, the pebble accretion model could also represent a good alternative scenario for planets formation. The contribution from pebbles to the mass increase of embryos would accordingly decrease the number of planetesimal-embryos impacts, and thus limit the crustal stripping during accretion, unless embryo-embryo impacts can also cause significant preferential crustal stripping. One should however note that recent studies have enlighten that terrestrial planet formation via pebble accretion seems to violate the observed nucleosynthetic anomalies dichotomy between the inner and outer solar system \citep[e.g.][]{burkhardt2021terrestrial, mah2022effects}. This argument is however still debated. For instance, \cite{brasser2020partitioning} already proposed that there was a pressure maximum in the disk around Jupiter formation location that may have limited inward drift of outer solar system pebbles, unabling for the preservation of early solar system dichotomy.  \newline

\newgeometry{left=1.5cm,bottom=1.5cm}

\begin{landscape}
\begin{table}
\centering
\begin{tabular}{|c||c|c|l|}
\hline
Couple & $\epsilon$ (a) & $\epsilon$ (b) & Summary\\

\hline

Sm-Nd & {\small $0.03\pm0.02$} & {\small $0.054\pm0.016$} & \begin{tabular}[c]{@{}l@{}}{\scriptsize The 20ppm-excess systematically measured in terrestrial samples compared} \\{\scriptsize to chondritic values may be fully explained by collisional erosion of the crust}\\~{\scriptsize during Earth formation in the context of Grand Tack with a late Moon-forming}\\~{\scriptsize impact. D$_{Sm}=$0.045 and D$_{Nd}=$0.031. }\end{tabular}                                                                                                                                          \\
\hline
Lu-Hf & {\small $0.09\pm0.04$} & {\small $0.16\pm0.05$} & {\scriptsize Same conclusion as for Sm-Nd. D$_{Lu}=$0.120 and D$_{Hf}=$0.035.}                                                                                                                                                                                                                                                                                                                                                                                                                                    \\
\hline
Mg-Si & {\small $0.007\pm0.003$} & {\small$0.01\pm0.003$} & \begin{tabular}[c]{@{}l@{}}{\scriptsize Poorly affected by crustal stripping during accretion history. This was expected} \\{\scriptsize for Major elements since the poorly partition into the melt when crust-forming}\\~{\scriptsize event occurs. D$_{Mg}=$3.98 and D$_{Si}=$0.90.}\end{tabular}                                                                                                                                                                                                            \\
\hline
Mn-Na & {\small $0.076\pm0.024$} & {\small $0.113\pm0.026$} & \begin{tabular}[c]{@{}l@{}}{\scriptsize Mn in not affected by the preferential erosion of the crust, as expected from its}\\~{\scriptsize middly incompatible property, however some Na is lost. This raises a paradigm}\\~{\scriptsize since Mn/Na is supposed to be chondritic in Earth. D$_{Mn}=$0.93 and D$_{Na}=$0.13.}\end{tabular}                                                                                                                                                                        \\
\hline
Rb-Sr & {\small $-0.16\pm0.06$} & {\small $-0.26\pm0.06$} & \begin{tabular}[c]{@{}l@{}}{\scriptsize The Rb/Sr ratio is expected to be fractionated compared to chondrites due to} \\{\scriptsize some volatilization post-nebula processes. 20\% of the total expected fractionation}\\~{\scriptsize from current estimates might be due to collisional erosion considering our results.}\\~{\scriptsize D$_{Rb}=$0.00001 and D$_{Sr}=$0.025.}\end{tabular}                                                                                                                     \\
\hline
Th-U  & {\small $-0.008\pm5.10^{-4}$} & {\small $-0.009\pm6.10^{-4}$} & \begin{tabular}[c]{@{}l@{}}{\scriptsize The Th/U ratio seems to undergo a very low fractionation. However, considering}\\~{\scriptsize uncertainties on measurements we cannot conclude on the values here.}\\~{\scriptsize In terms of absolute concentrations in the BSE for Th and U, less than half of}\\~{\scriptsize them is lost during accretion. This still be consistent with the current estimates}\\~{\scriptsize provided either by heat fluxes or geoneutrinos measurements.} \\~{\scriptsize D$_{Th}=$0.001 and D$_{U}=$0.0011.}\end{tabular} \\
\hline

\end{tabular}
	\caption{Table of the $\epsilon$ average values in the BSP after the growth of the embryos considered as good Earth analogs in a grand tack accretion scenario\citep{jacobson2014lunar} along with their associated to their standard deviation for all couples of elements studied here. $\epsilon$ (a) refers to the average $\epsilon$ associated to all relevant Earth analogs while the $\epsilon$ (b) column refers only to Earth analogs recording a last giant impact after 50 Myr. The last column of this table presents a quick summary of the conclusions we can propose from these values. The partitioning coefficients of the different elements are also given (D$_M$). The values are taken from \cite{workman2005major}.} 
	\label{tab:conclusion}
\end{table}
\end{landscape}
\restoregeometry

\section*{Acknowledgement}
J.S. acknowledges support from the French National Research Agency (ANR project VolTerre, grant no. ANR- 14-CE33-0017-01) And Institut Universitaire de France.
Parts of this work were supported by the UnivEarthS Labex programme at Sorbonne Paris Cit\'e (ANR-10-LABX-0023 and ANR-11- IDEX-0005-02).


\bibliography{mybibfile}
\newpage

\section*{Appendices}
	\label{sec:annexes}

In this appendice we detail the erosion model (EROD) that we developed by: (1) the description of the SFD used to describe the planetesimals impacting an embryo; (2) the cratering model used for the estimates of the mass transfer. Part of this model has been developed and described in \citep[][]{allibert2021quantitative}. 
	
\subsection{Planetesimals size-frequency distribution (SFD) imposed}
	
N-body simulations need high computational requirements and cannot include a realistic amount of planetesimals. Planetesimals are test particles meant to represent for an entire population of smaller bodies whom total mass is equal to the planetesimal mass. Accordingly, for the mass transfer estimates required by our model we distribute the mass of a single planetesimal impactor into a size-frequency distribution (i.e. each impact from a planetesimal in the N-body simulations is assumed to be a serie of impacts with impactors distributed on a size-frequency distribution (SFD)). That serie of impactors is assumed to fall onto the target at the same time and to be coming from all directions.

	The differential size distribution of impactors follows a power law   :
	
	\begin{equation}
	    \frac{dN_{sfd}(>r)}{dr}=-Kr^{-\alpha},
	\end{equation}
	with $\alpha$ and K standing for positive constants. $N_{sfd}(>r)$ is the cumulative size distribution (the number of bodies with a radius larger than $r$). 
	
	The minimum and maximum masses that a body can take in the distribution have to be previously fixed. They are arbitrarily chosen as 8m and 800km respectively \citep[e.g.][]{weidenschilling2011initial, johansen2014multifaceted}. The number of bodies in each mass bin is evaluated using the cumulative mass distribution, rounded to an integer number by using a random number generator. The traduces the fact that the biggest possible bodies in the simulations are more rare. The average density of embryos is assumed to be $5200\ kg/m^3$. The crust density is set as $2900\ kg/m^3$ while impactors density (chondritic material) is assumed to be $2600\ kg/m^3$. The impact velocity can be expressed as $V_{imp}=\sqrt{V_{\inf}^2+v_{esc}^2}$ with $V_{\inf}$ the velocity of the impactor at infinity (beyond the gravitational attraction of embryos) and $v_{esc}$ the escape velocity of the two-bodies system. We assume that $V_{\inf}$ is the same for all impactors in a given distribution. $V_{\inf}$ is deduced from N-body simulations outcomes. The radii of target and impactors are estimated according to their average densities and their respective masses.

	The slope of the SFD ($-\alpha$) is set at $-3.5$ which corresponds to the actual value of the distribution of bodies in the asteroid belt \citep[e.g.][]{hellyer1970fragmentation, paolicchi1994rushing, williams1994size, tanaka1996steady, bottke2005fossilized}.  

\subsection{Cratering model for E-P impacts}

   The mass transfer (accretion and erosion) during a single impact effect is deduced from analytical laws constructed as a function of the impact parameters (velocity at encounter, the impactor and target respective densities and the impactor to target mass ratio)  \citep{holsapple2007crater, shuvalov2009atmospheric, svetsov2011cratering}.
	   			
Embryos are assumed to be differentiated into a core, a mantle and a crust. The model assumes that all impactors from the size-frequency distribution come from all directions. Accordingly, the ejected mass is assumed to be of crustal composition, if the total ejected mass (i.e. after considering the cumulative effects of all impacts from a given size-frequency distribution) is lower than the mass of the crust itself. This then takes into account the fact that overlapping between the different craters may occur (in opposition of a method using instead the crater depth only as a parameter to decipher from crust-mantle composition in the ejecta). For greater ejected masses, the remnant mass is assumed to have mantle composition. All planetesimals are assumed to be chondritic in their bulk composition: accreted mass has then a chondritic composition in the model. \newline

\end{document}